\documentclass[12pt,twoside]{article}
\usepackage[T1]{fontenc}
\usepackage[latin1]{inputenc}
\usepackage{times}
\usepackage{graphicx}
\usepackage{a4wide}
\usepackage{listings}
 \usepackage{amsmath}

\begin{document}

\section*{Meson and di-electron production with HADES\footnote{Invited plenary talk
    at the 10th International Workshop On Meson Production, Properties
    And Interaction (MESON 2008) 6-10 Jun 2008, Cracow, Poland }}

\thispagestyle{empty}

\begin{raggedright}

\markboth{I.Fr\"ohlich {\it et al.}}
{Meson and di-electron production with HADES}

I.~Fr\"{o}hlich$^{7}$, G.~Agakishiev$^{8}$, C.~Agodi$^{1}$,
A.~Balanda$^{3,e}$, G.~Bellia$^{1,a}$, D.~Belver$^{15}$, A.~Belyaev$^{6}$,
A.~Blanco$^{2}$, M.~B\"{o}hmer$^{11}$, J.~L.~Boyard$^{13}$
% ...

\begin{center} (HADES collaboration)
\end{center}

{$^{1}$}Istituto Nazionale di Fisica Nucleare - Laboratori Nazionali del Sud, 95125~Catania, Italy\\
{$^{2}$LIP-Laborat\'{o}rio de Instrumenta\c{c}\~{a}o e F\'{\i}sica Experimental de Part\'{\i}culas, 3004-516~Coimbra, Portugal}\\
{$^{3}$Smoluchowski Institute of Physics, Jagiellonian University of Cracow, 30-059~Krak\'{o}w, Poland}\\
{$^{4}$Gesellschaft f\"{u}r Schwerionenforschung mbH, 64291~Darmstadt, Germany}\\
{$^{5}$Institut f\"{u}r Strahlenphysik, Forschungszentrum Dresden-Rossendorf, 01314~Dresden, Germany}\\
{$^{6}$Joint Institute of Nuclear Research, 141980~Dubna, Russia}\\
{$^{7}$Institut f\"{u}r Kernphysik, Johann Wolfgang Goethe-Universit\"{a}t, 60438 ~Frankfurt, Germany}\\
{$^{8}$II.Physikalisches Institut, Justus Liebig Universit\"{a}t Giessen, 35392~Giessen, Germany}\\
{$^{9}$Istituto Nazionale di Fisica Nucleare, Sezione di Milano, 20133~Milano, Italy}\\
{$^{10}$Institute for Nuclear Research, Russian Academy of Science, 117312~Moscow, Russia}\\
{$^{11}$Physik Department E12, Technische Universit\"{a}t M\"{u}nchen, 85748~M\"{u}nchen, Germany}\\
{$^{12}$Department of Physics, University of Cyprus, 1678~Nicosia, Cyprus}\\
{$^{13}$Institut de Physique Nucl\'{e}aire d'Orsay, CNRS/IN2P3, 91406~Orsay Cedex, France}\\
{$^{14}$Nuclear Physics Institute, Academy of Sciences of Czech Republic, 25068~Rez, Czech Republic}\\
{$^{15}$Departamento de F\'{\i}sica de Part\'{\i}culas, University of Santiago de Compostela, 15782~Santiago de Compostela, Spain}\\
% ...
$^{a}${Also at Dipartimento di Fisica e Astronomia, Universit\`{a} di Catania, 95125~Catania, Italy}\\
% ...
$^{c}${Also at Technische Universit\"{a}t Dresden, 01062~Dresden, Germany}\\
% ...
$^{e}${Also at Panstwowa Wyzsza Szkola Zawodowa, 33-300~Nowy Sacz, Poland}

\end{raggedright}

\begin{center}
\textbf{Abstract}
\end{center}
  The HADES experiment, installed at GSI, Darmstadt, measures di-electron
  production in $A+A$, $p/\pi+N$ and $p/\pi+A$ collisions. Here, the $\pi^0$ and
  $\eta$ Dalitz decays have been reconstructed in the exclusive $p+p$ reaction
  at 2.2~GeV to form a reference cocktail for long-lived di-electron sources.
  In the C+C reaction at 1 and 2 GeV/u, these long-lived sources have been
  subtracted from the measured inclusive $e^+e^-$ yield to exhibit the signal
  from the early phase of the collision.  The results suggest that resonances
  play an important role in dense nuclear matter. 

\section{Introduction}  

One of the main subjects in physics is how hadrons are generated by the strong
interaction and how hadron-hadron forces can be described. Here, effective
models have to be employed to calculate properties of hadrons and their
interactions at different environmental conditions. In the vacuum, e.g.,
One-Boson exchange (OBE) models describe the production and (de-)excitation of
hadrons taking intermediate resonances into account in a coherent approach.
On the other hand, the evolution of hadronic matter (like in heavy ion
reactions) is usually calculated by transport models where different processes
are factorized. The question of in-medium modifications of hadronic spectral
functions arises additional complexity. This investigation has in particular
raised by the question if chiral symmetry restoration appears in hot/dense
matter which could lead to dropping masses of hadrons~\cite{br_scaling}.
Therefore, the creation of nuclear matter under extreme conditions in the
laboratory (possible only in A+A reactions) is a unique tool to study and
understand hadronic properties.

In this context, the emission of virtual photons ($\gamma^*$), decaying into
di-leptons ($e^+e^-$ or $\mu^+\mu^-$) which do not undergo strong interaction,
has been proposed as an ideal probe for such studies.  Hence, this undistorted
``light'' from the early phase of the collision allows the study what could be
the nature of hadronic matter over the whole nuclear phase region (baryonic
matter, meson gas, quark-gluon phase). Virtual photons, moreover, couple via
virtual vector mesons ($\rho^0,\omega$) to the hadronic matter ({\bf v}ector
{\bf m}eson {\bf d}ominance model). A possible change of vector meson
properties (e.g.  dropping mass, larger width) can be studied by separating
the di-lepton cocktail into two parts: The radiation from short-lived
resonances (as discussed above) is extracted by subtracting the long-lived
components (i.e. electromagnetic de-excitation of hadrons after the freeze-out)
from the measured $e^+e^-$ spectrum.  

Indeed, it has been found that di-lepton spectra taken in heavy-ion
collisions~\cite{na45,na60,phenix} need additional short-lived sources in the
invariant-mass region 0.2~GeV/c$^{2}\leq M_{ee} \leq$0.6~GeV/c$^{2}$.  At
least in the SPS energy range this feature has been related to modifications
of the $\rho$-meson spectral function in the hadronic medium~\cite{hees_renk}.

However, for a complete understanding of hadronic properties it is important
to measure not only the hot, but also the dense nuclear matter.  At beam
energies of 1-2~GeV/u, which correspond to moderate densities (2-3 $\rho_0$)
and temperatures (60-80~MeV), the production of mesons is dominated by
multi-step processes with excitation of intermediate baryonic resonances like
$\Delta^{+,0}\to N\pi^{0}\to N\gamma e^{+}e^{-}$ ($\pi$-Dalitz),
$N^{*}(1535)\to N\eta\to N\gamma e^{+}e^{-}$ ($\eta$-Dalitz) and the Dalitz
decays of baryonic $\Delta, N^*$ resonances in $N(\omega,\rho) \to N\gamma^*
\to N e^{+}e^{-}$.  This means, after removing the long-lived components (the
Dalitz decays $\pi^0,\eta$ have a known form factor~\cite{landsberg}) the
remaining di-lepton strength represents the electromagnetic response of
resonant matter, which is even on the elementary basis completely unknown. For
example, the Dalitz decays of the baryonic resonances (including the
$\Delta$(1232)) have not been measured. In the overall picture, these
contributions are additional exchange graphs in the virtual bremsstrahlung
process $NN\to NN\gamma^*$. One of the the questions recently addressed by OBE
models is how the resonance contributions have to be treated among with the
bremsstrahlung in coherent calculations.  In this context it should be pointed
out that the large pair yields in the region dominated by resonance decays
(0.2-0.6~GeV/c$^2$) observed by the pioneering DLS experiment~\cite{dls} in
$C+C$ and $Ca+Ca$ collisions at 1~GeV/u remain to be explained
satisfactorily~\cite{dls_collection}.  Additional contributions from the
bremsstrahlung graphs could possibly explain the missing sources but a debate
on this is still ongoing~\cite{kaptari,mosel}.  The general conclusion is,
however, that experimentally a strong isospin dependence should be visible in
the mass-dependent ratio $M^{pp}_{ee}/M^{pn}_{ee}$.

To conclude this introduction, for an understanding of di-lepton spectra and
resolving medium effects from the vacuum spectral functions it is crucial to
fully describe the different (maybe coherent) sources in the heavy ion
cocktail. 

\section{The HADES program}

The High-Acceptance DiElectron Spectrometer HADES at GSI, Darmstadt, is
presently the only di-lepton experiment operational in the SIS energy regime of
\mbox{1-2~GeV/u}, succeeding the DLS experiment.  To summarize the setup
at this point, HADES~\cite{schicker} is a magnetic spectrometer,
consisting of up to 4 planes of Mini Drift Chambers (MDC) with a toroidal
magnetic field.  Particle identification is based on momentum and
time-of-flight measurements.  In addition, a Ring Imaging Cherenkov detector
(RICH) and an electromagnetic Pre-Shower detector provide electron
identification capabilities.  To increase the acquired pair statistics,
besides a charged-particle multiplicity trigger (LVL1), an online electron
identification (LVL2) has been operated as part of the two-level trigger
system~\cite{traxler}.

In order to address the open questions outlined above, one of the goals of the
HADES detector system is to measure the di-electron properties not only in heavy
ion collisions but in elementary reactions as well.  Consequently, the HADES
program spans from $p+p$, $\pi+ p$ to $\pi+ A$ and $A+A$ collisions.  First
data has been taken in $C+C$ collisions~\cite{prl,plb} at 1 and 2~GeV/u,
$Ar+KCl$ collisions at 1.756~GeV/u, and interactions of $d+p$ at 1.25~GeV/u,
and $p+p$ at 1.25, 2.2 and 3.5~GeV.

The idea of these measurements is to fully disentangle and describe the
elementary sources which are needed to understand the heavy ion case.
Here, $p+p$ and $d+p$ measurements, which are described in detail in a
separate contribution~\cite{tetyana}, allow
to study the $\Delta$ Dalitz decay which is the main di-electron contribution
above the $\pi^0$ mass at these energies.  In the $p+p$ reaction at 2.2~GeV
the $\pi^0/\eta$ Dalitz decays can be disentangled by exclusive studies,
which will be shown in the first part of this contribution.  This
measurement works also as a reference reaction for the long-lived
components of the $C+C$ heavy ion data.  The $p+p$ reaction at 3.5~GeV
was mainly done to measure  the free $\omega$ line shape and to understand
possible contributions of higher resonances.  In $C+C$
collisions we have extracted the radiation by short-lived resonances,
which is discussed in this contribution.  In addition, vector meson
production has been measured in the $Ar+KCl$ (described separately as
well~\cite{filip}).
As an outlook, future perspectives of the HADES experiment are given.

\section{Exclusive meson production in the $p+p$ reaction at 2.2~GeV}

One aim of the $p+p$ experiment done at a kinetic beam energy of 2.2 GeV was
to have a reference reaction for the $C+C$ reaction at 2~GeV/u and study the
Dalitz decays $pp \to pp \pi^0 \to pp \gamma e^+e^-$ and $pp \to pp \eta \to
pp \gamma e^+e^-$ independently by a selection on the $pp$ missing mass. This
allows to exclude uncertainties for this part of the cocktail. But before
conclusions on these decays are drawn the parameters used by simulations have
to be confirmed for the acceptance correction of the di-electron spectra.
Available data in $N+N$ collisions at this energy suffers from the following.
In the $\eta$ case, the total cross section has been determined by a
fit~\cite{phi_omega} taking into account data points with large errors.  The
$pp \to pp \pi^0$ total cross section is known~\cite{teis}, but more resonances
with non-isotropic behavior are involved.  For instance, the pion scattering
angle distribution $\theta^\Delta_{\pi}$ in the $NN \to N \Delta \to NN \pi$
reaction has been found by bubble chamber experiments to be in disagreement to
one-pion exchange~\cite{wicklund}.

Therefore, the $pp \to pp M$ reactions (where $M$ is either a $\pi^0$
or $\eta$) have been compared to simulations for which we usually
employ the Pluto event generator~\cite{pluto}.  The normalization of
these distributions has been done by comparing the yield of measured
mesons $N^M$ to the yield of $pp$ elastic events $N^{el}$, and a model
dependent Pluto simulation with the known $pp$ elastic cross section
and angular distribution~\cite{said,pp_el}. To take detector effects
into account, each measured event is corrected first for 
single-track detector and reconstruction inefficiencies. The simulated
events, on the other hand, are filtered through an acceptance filter
(a matrix $A_{\pm}(p,\theta,\phi)$ for each track) and smeared in
momentum using a parameterized resolution.  After this
procedure, the resulting spectra can be compared to the simulations.
Uncertainties due to different production mechanisms are evaluated by
varying the different parameters.


\begin{figure}[t]
  \begin{center}
    \parbox[c]{0.48\textwidth}{
      \centering
      \includegraphics[width=0.48\textwidth]{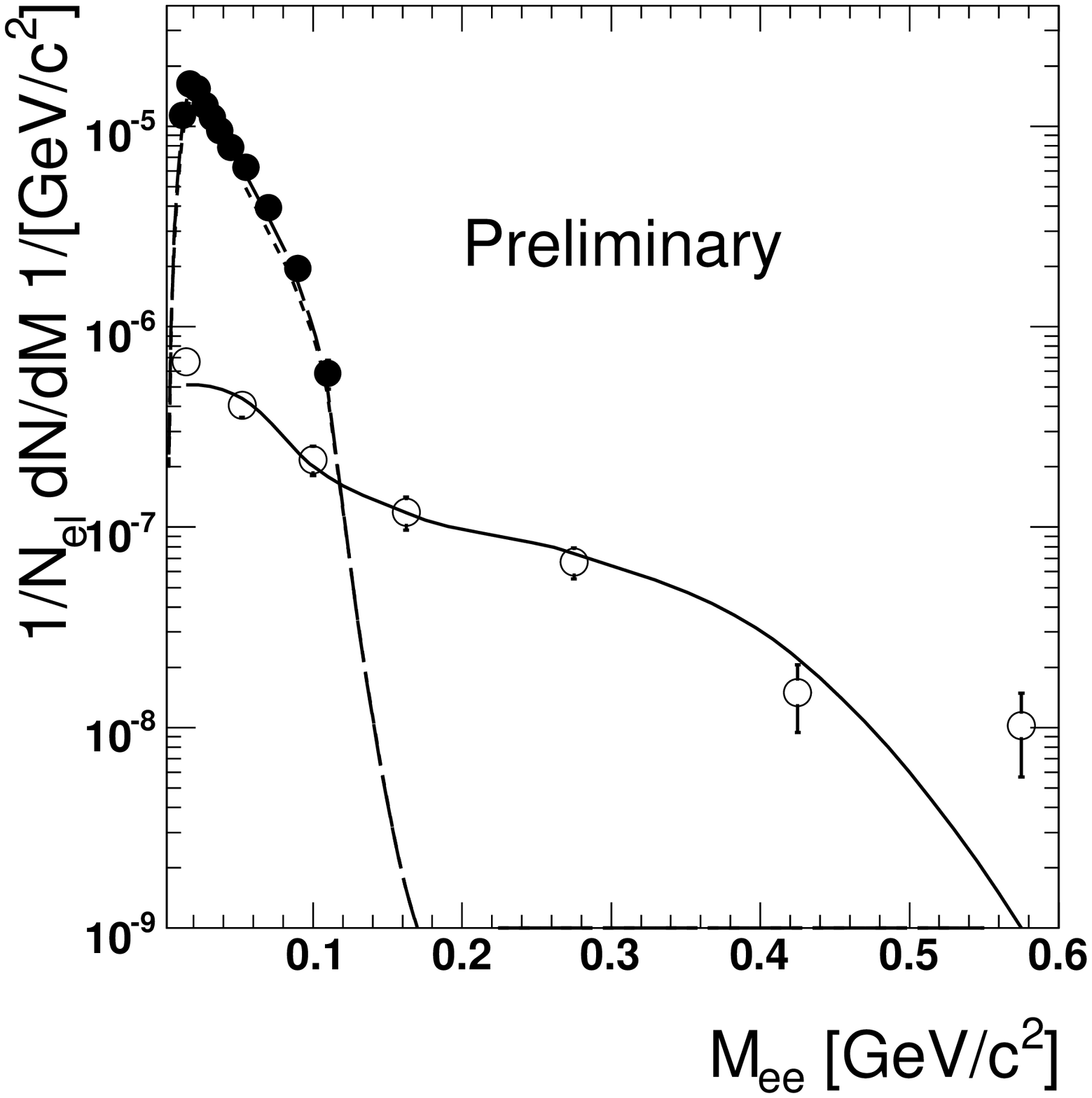}
    }
    \parbox[c]{0.48\textwidth}{
      \centering
      \includegraphics[width=0.48\textwidth]{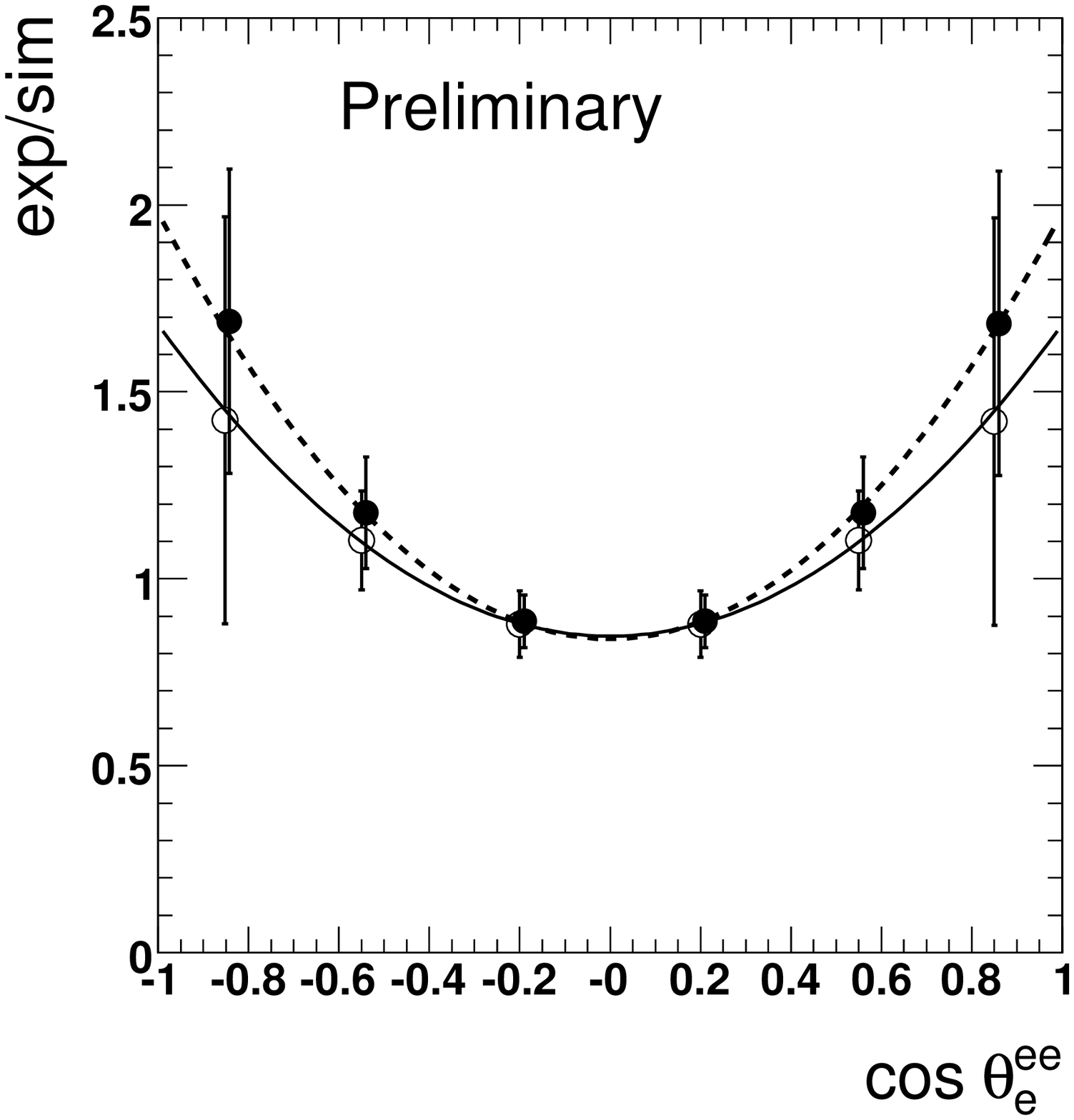}
    }
    \caption{ {\bf Left:} Di-electron mass spectra for the $\pi^0$ (full symbols)
      and $\eta$ (open symbols) Dalitz decay (corrected
      for efficiency and normalized to the number of $pp$ elastic
      events). The solid curve shows the simulated $\eta$ Dalitz decay
      (acceptance taken into account), whereas the dashed curves represent the
      $\pi^0$ Dalitz decay (long-dashed with $\Delta$ decay angle from
      Ref.\protect~\cite{wicklund}, short-dashed with one-pion exchange model.)
      {\bf Right:} Helicity angle distribution among with a fitted function
      ($a \cdot (1 + b\cdot cos^2 \theta_e^{ee})$) for $\pi^0$ (full symbols
      and dashed line) and $\eta$ (open symbols and solid line). }
    \label{minv}
  \end{center}
\end{figure}

We have studied the reactions $pp \to pp \pi^0$ and $pp \to pp \eta \to pp
\pi^+ \pi^- \pi^0$. In both cases, the neutral $\pi^0$ (not detected in the
setup) has been selected by the missing mass method. For the $\eta$ simulation
the polar angle distributions and resonance model from DISTO~\cite{disto},
obtained at a similar beam energy (the ratio $\frac{\sigma (pp \to pN^*1535
  \to pp \eta)}{\sigma (pp \to pp \eta) }$ is 1.38), and the $\eta$ decay
properties from Crystal Barrel~\cite{cbarrel1} have been used.  The preliminary
results $\sigma_{pp \to pp \eta}=70 \pm 8$mb is consistent with the
parametrizations.  In the $\pi^0$ case the resonance model~\cite{teis} has
been adapted and included into Pluto.  Unlike the $\eta$ case, angular
distributions (for the production of resonances as well as for their decay)
play an much larger role.  For the $\Delta$ and $N^*$ production the strong
forward-backward peaking of the polar angle distribution has been taken into
account~\cite{cite_13_dmitriev,aichelin}.  For the distribution of
$\theta^\Delta_{\pi}$ the measured coefficient~\cite{wicklund} was used. In
addition, the influence of applying a pure one-pion exchange model has been
evaluated. Our preliminary data suggests that we need additional resonance
contributions to describe the data~\cite{wisnia}. Here, we extracted
$\sigma_{pp \to pp \pi^0}=4.17\pm 0.05^{stat} \pm 0.48^{norm}$mb which is
slightly above the model~\cite{teis}.

In the following, we used these production parameters for a di-electron
cocktail simulation with $\eta/\pi^0$ components only to compare to our
exclusive data which we obtained as follows.  Opposite-sign reactions $pp \to
pp e^+e^-$, as well as like-sign $pp \to pp e^+e^+$ and $pp \to pp e^-e^-$
events were formed and subjected to common selection criteria (e.g., an
opening-angle cut of $\theta_{ee}>4^{\circ}$). In addition, a selection on the
Dalitz decays $M \to e^+e^-\gamma$ has been done by identifying a missing
$\gamma$.  From the reconstructed distributions first the combinatorial
background (CB) of uncorrelated pairs was calculated bin by bin as $N_{CB} =
2\sqrt{N^{++}N^{--}}$. Then the $e^+e^-$ pairs have been selected as a
function of the $pp$ missing mass. The $\pi^0$ and $\eta$ contribution has
been evaluated by fitting the signal plus background. By repeating this method
on different $e^+e^-$ mass slices, the $e^+e^-$ invariant spectrum can be
obtained.  Fig.~\ref{minv} (left part) shows the efficiency corrected data for
the $\pi^0$ and $\eta$ Dalitz decay, normalized to the number of elastic
scattered $pp$ events, and compared to the Pluto simulation. Possible effects
of the $\Delta$ decay angle are small. As a conclusion it can be drawn that the
$e^+e^-$ invariant mass of the $\pi^0$ and $\eta$ Dalitz decays are well
understood and the chosen production parameters match the data points.

In addition to the invariant mass the virtual photon (decaying into 2 stable
particles) has 5 more degrees of freedom~\cite{qnp}: One of them, the helicity
angle $\theta^{ee}_e$, is related to the longitudinal polarization of the
virtual photon. From QED considerations~\cite{brat} this distribution has to
be $1+\cos^{2}\theta^{ee}_e$. In Fig.~\ref{minv} (right
part) the result (after a model-dependent correction) has been fitted with $a
\cdot (1 + b \cdot cos^2 \theta^e_{ee})$ which gave $b=1.35 \pm 0.44$ and
$b=0.98 \pm 0.48$ for the $\pi^0$ and $\eta$, respectively, in agreement to
the calculation~\cite{brat}.

\section{Di-electron production in $C+C$ collisions at 1 and 2~GeV/u}

\begin{figure}[t]
  \begin{center}
    \parbox[c]{0.45\textwidth}{
      \centering
      \includegraphics[width=0.45\textwidth]{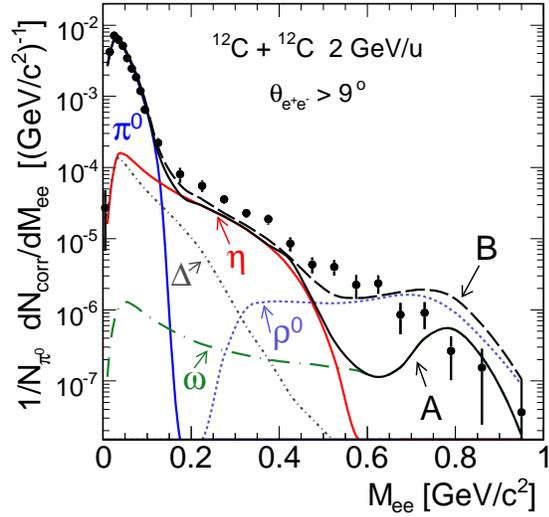}
    }
    \parbox[c]{0.53\textwidth}{
      \caption{
        HADES Di-electron mass spectrum (C+C, 2~GeV/u) compared to
        cocktail A (long-lived) and B (including short-lived sources). }
      \label{fig_plb}
    }    
  \end{center}
\end{figure}

HADES has measured di-electron spectra obtained in $C+C$
collisions~\cite{prl,plb} at two different beam energies (1 and 2~GeV/u). The
first data set allows to be compared to that of the former DLS
experiment~\cite{dls}. In order to address the open questions raised by the DLS
experiment we follow the usual strategy and subtract the long-lived components.

In the pair analysis (details of the data analysis can be found
in~\cite{prl,plb}), opposite-sign pairs ($e^+e^-$), as well as like-sign pairs 
$(e^+e^+$ and $e^-e^-$) were formed and an opening-angle cut of
$\theta_{ee}>9^{\circ}$ has been used.  From the reconstructed
distributions the combinatorial background (CB) was formed and
subtracted as in the previous case.  For masses $M_{ee}>0.2$
GeV/$c^2$, where statistics is small, the CB was obtained by an
event-mixing procedure. Detector and reconstruction efficiencies as
well as the acceptance were taken into account as described
above. Here, the normalization was done by the number of charged pions
$N_{\pi^0} = 1/2 (N_{\pi^+} + N_{\pi^-})$, as measured also in HADES
and extrapolated to the full solid angle~\cite{pionpaper}.  The
obtained pion multiplicity per participant nucleon, i.e.
$M_{\pi}/A_{part} = 0.061 \pm 0.009$ for 1~GeV/u and $M_{\pi}/A_{part}
= 0.153 \pm 0.023$ for 1~GeV/u, agrees well with previous measurements
of charged and neutral pions~\cite{averbeck}.

\begin{figure}[t]
  \begin{center}
    \parbox[c]{0.49\textwidth}{
      \centering
      \includegraphics[width=0.49\textwidth]{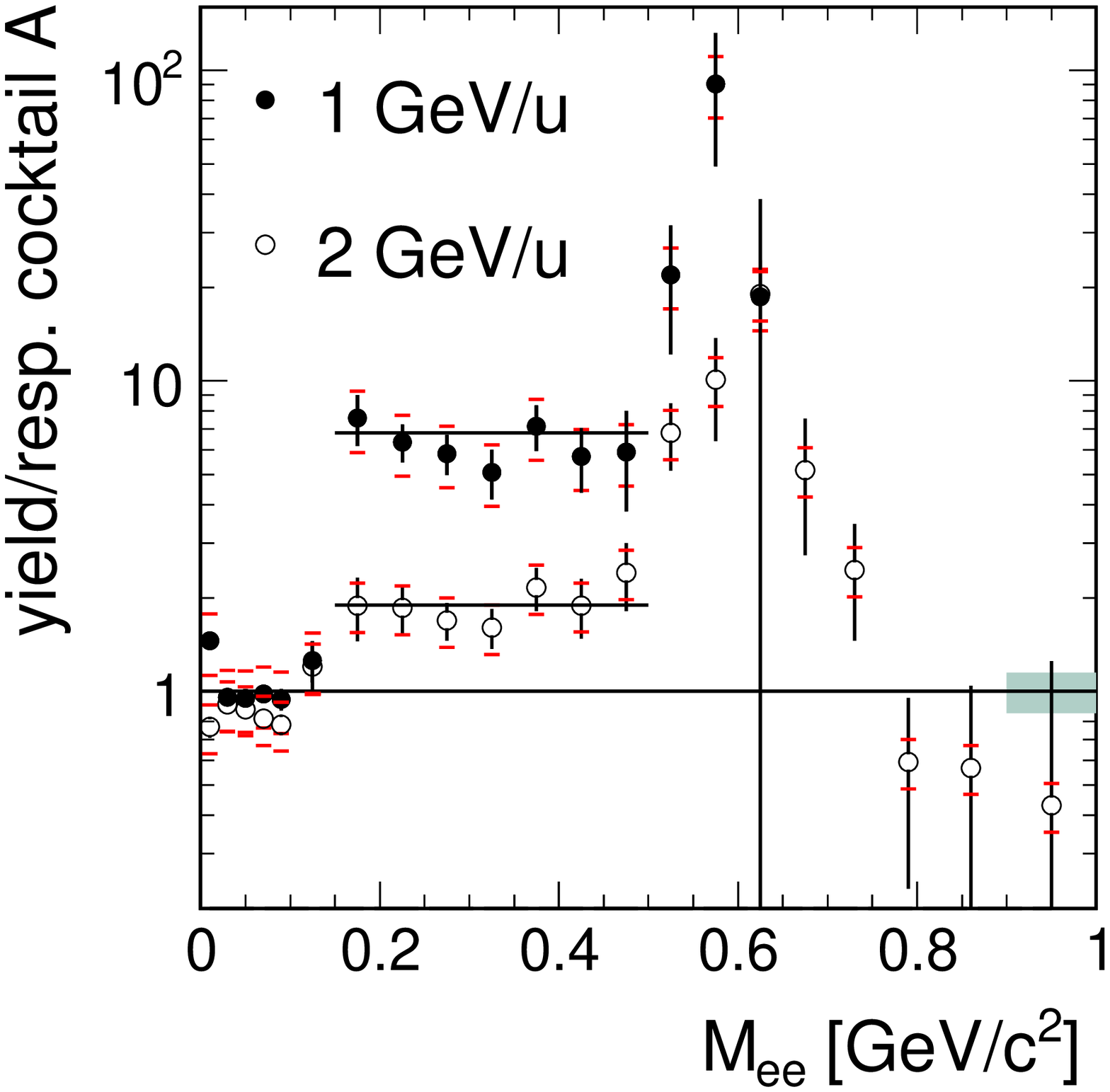}
    }
    \parbox[c]{0.49\textwidth}{
      \vspace{-0.3cm}
      \includegraphics[width=0.49\textwidth]{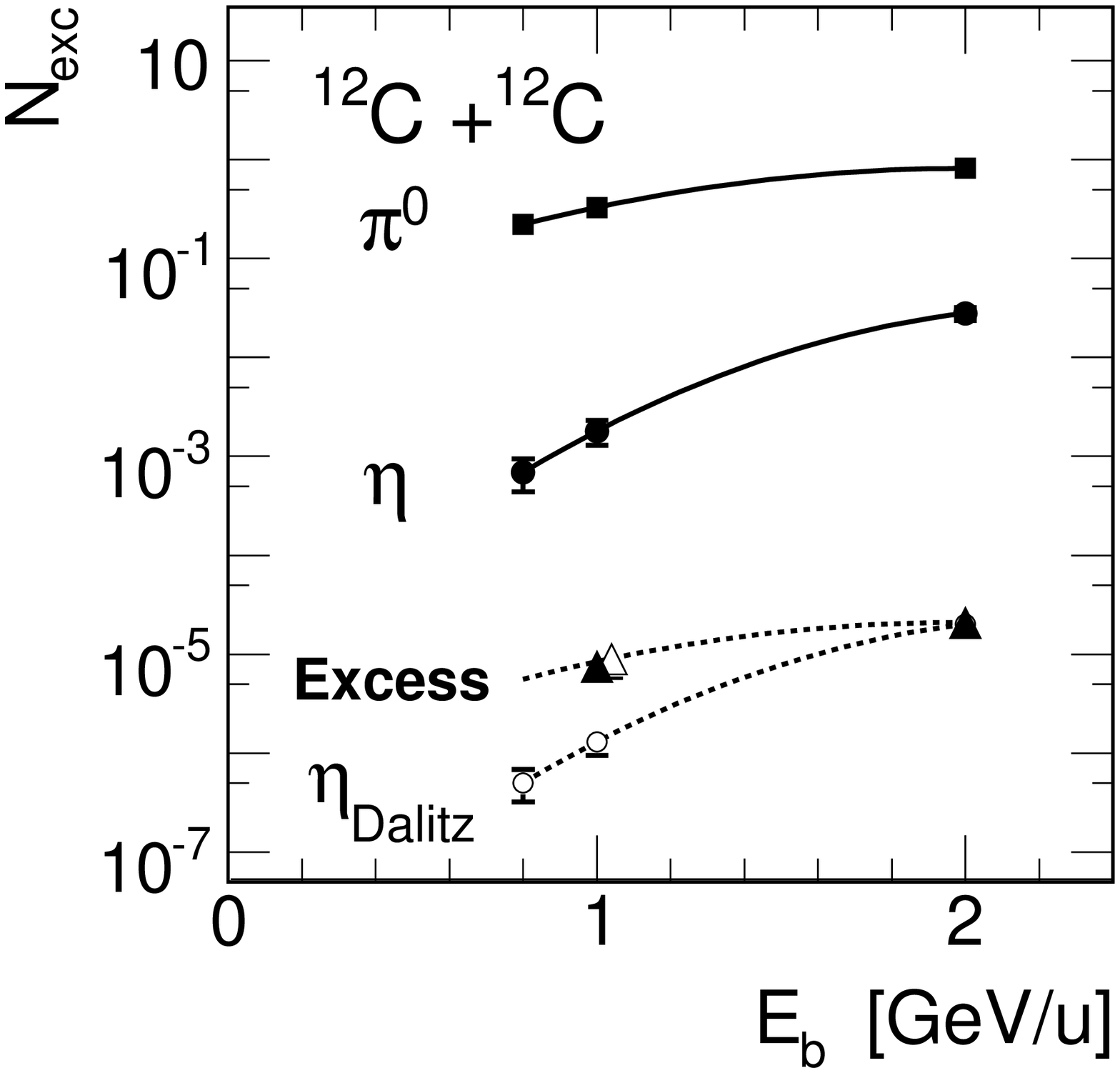}
    }
    \vspace*{8pt}
    \caption{  {\bf
        Left:} Experimental yield divided by cocktail A for 1~GeV/u (full
      symbols) and 2~GeV/u (open symbols) data.  In addition, the ratio of
      cocktail B and A for 1 GeV/u data is indicated (dashed line).  {\bf
      Right:} Excess over cocktail A compared to the measured multiplicities
      of $\eta$ and $\pi^0$ as a function of bombarding energy} 
    \label{fig_plb2}
  \end{center}
\end{figure}

In Fig.~\ref{fig_plb} the measured $e^+e^-$ invariant mass (2~GeV/u data) is
compared to a pair cocktail calculated from free $\pi^0 \rightarrow \gamma
e^{+}e^{-}$, $\eta \rightarrow \gamma e^{+}e^{-}$, $\omega \rightarrow
\pi^{0}e^{+}e^{-}$, and $\omega \rightarrow e^{+}e^{-}$ meson decays only
(cocktail A), simulated with the Pluto package~\cite{pluto}, representing the
radiation from long-lived mesons. As expected, experimental data and
\mbox{cocktail A} agree in the $\pi^0$ Dalitz region. In the mass region
$M_{ee}>0.15$ $GeV/c^{2}$ the cocktail strongly underestimates the measured
pair yield.  This is not surprising, since one expects additional
contributions from short-lived resonances. The cocktail B (Fig.~\ref{fig_plb},
long-dashed line) considering additional $\Delta(1232)$ and $\rho$ components
by a simple estimation is still not sufficient to describe the signal.

As pointed out above, by subtracting the long-lived sources, the signal from
the early phase of the dense matter should become visible.  This is shown for
both measured beam energies in Fig.~\ref{fig_plb2} (left). A structure in the
vector meson region at 0.6 GeV can be clearly seen and may be attributed to a
virtual $\rho^0$ excitation.  In addition, a large excess over the cocktail A
in the mass region $ 0.15\: GeV/c^{2} \leq M_{ee} \leq 0.5\: GeV/c^{2}$ can be
found.  The assumption that this enhancement is caused by an underestimated
$\eta$ yield and/or not well understood efficiency of the detector is, as
demonstrated in the previous section, ruled out.  To further understand this
excess, it has been integrated and compared to the measured multiplicities of
$\eta$ and $\pi^0$ as a function of bombarding energy (Fig.~\ref{fig_plb2},
right part).  The result is that the excess scales with $\pi^0$ production.
This indicates that resonances, correlated via their main decay branch to pion
production, play a much larger role than usually supposed. It should be
notices at this point that the virtual bremsstrahlung process, described in a
recent OBE model~\cite{kaptari}, cannot be factorized into a resonance part and
the elastic term. In order to understand these processes and the electromagnetic
structure of the $\Delta$ resonance on an elementary basis, dedicated
experiments ($p+p$ and $d+p$ at 1.25~GeV/u) have been done by
HADES~\cite{tetyana}.

\section{Vector meson production in $p+p$ at 3.5~GeV}

\begin{figure}[t]
  \begin{center}
    \parbox[c]{0.47\textwidth}{
      \centering
      \includegraphics[width=0.47\textwidth]{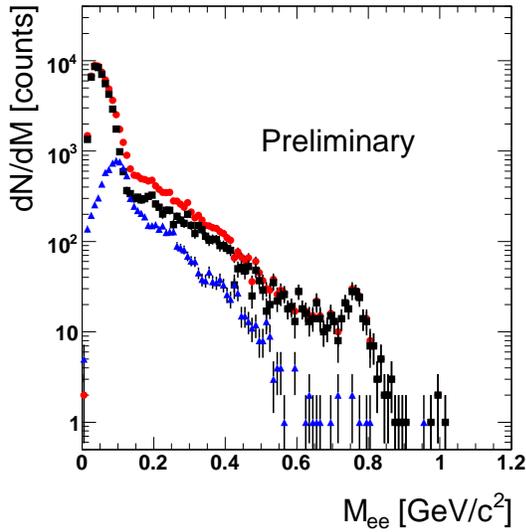}
    }
    \parbox[c]{0.51\textwidth}{
      \caption{
        HADES Di-electron mass spectrum (p+p, 3.5~GeV). 
        Circles: all pairs. Triangles: combinatorial background.
        Boxes: true signal. The data is not corrected for efficiency.
      }
      \label{fig_pp3.5}
    }
  \end{center}
\end{figure}

In the previous two sections we have demonstrated that the di-electron
contributions of $\pi^0$ and $\eta$ in elementary as well as in heavy ion
collisions is well understood. The additional yield seen above the eta Dalitz
cannot be explained from long-lived sources, suggesting that it comes from the
early phase of the collisions and seems to be the glowing of resonant matter.
But there are still some questions left region, which are (in part): 1.)  what
is the production mechanism of the $\omega$ (including resonances), and 2.)
what is the correct description of the free $\rho$ spectral function and how
does it change inside nuclear matter.  To give additional constraints on these
questions HADES has measured the di-electron production in the $p+p$ reaction
at 3.5~GeV. Fig.~\ref{fig_pp3.5} shows the (preliminary) invariant mass
spectrum without any efficiency correction. The $\omega$ peak is clearly seen.
Further exclusive studies can now, similar to the strategy already outlined,
further disentangle the di-electron cocktail in the vector meson region and
conclusions about the $\rho$ mass shape in vacuum and its coupling to
resonances can hopefully been drawn. Moreover, the $\omega$ line shape obtained
by this experiment can be compared to that one in cold nuclear matter. As past
experiments about this questions were not conclusive~\cite{metag,taps}, HADES will
take data using the $p+Nb$ collision in the near future at the same beam
energy of 3.5~GeV which allows for a direct comparison of the spectra.

\section{Future and outlook}

HADES will continue its program at its current place at SIS, and then move to
the upcoming FAIR
accelerator complex. 
This is one of the main reason for upgrading~\cite{rpc} the HADES detector and
its trigger and readout system~\cite{ieee} with the addition of the new RPC
detector (``Resistive Plate Chamber''). Since the currently used data
acquisition system was designed ten years ago, it was reasonable to reconsider
the whole concept and make use of new technologies. Furthermore, the large
data volumes, expected in experiments with heavy collision systems (Au+Au)
already at SIS18 and with higher energies at the new FAIR facility, require
bandwidths which cannot be achieved by the current system.

Here, HADES will continue the systematic investigation, successfully started
at SIS, up to kinetic beam energies of 8~GeV/u at FAIR and step, as there is
no dilepton data between 2 and 40~GeV/u so far, into complete ``terra
incognita''.

\section*{Acknowledgments}
\renewcommand{\baselinestretch}{0.9}
\footnotesize
The collaboration gratefully acknowledges the
support by BMBF grants 06MT238, 06GI146I,
06F-140, and 06DR135 (Germany), by the DFG
cluster of excellence Origin and Structure of
the Universe (www.universe-cluster.de), by GSI
(TM-FR1,GI/ME3,OF/STR), by grants GA CR
202/00/1668 and MSMT LC7050 (Czech Re-
public), by grant KBN 1P03B 056 29 (Poland),
by INFN (Italy), by CNRS/IN2P3 (France), by
grants MCYT FPA2000-2041-C02-02 and XUGA
PGID T02PXIC20605PN (Spain), by grant UCY-
10.3.11.12 (Cyprus), by INTAS grant 03-51-3208
and by EU contract RII3-CT-2004-506078.

\renewcommand{\baselinestretch}{1}

\end{document}